# On the Status of the Measurement Problem:
## Recalling the Relativistic Transactional Interpretation

R. E. Kastner
16 December 2017
Foundations of Physics Group, University of Maryland

ABSTRACT. In view of a resurgence of concern about the measurement problem, it is pointed out that the Relativistic Transactional Interpretation (RTI) remedies issues previously considered as drawbacks or refutations of the original Transactional Interpretation (TI). Specifically, once one takes into account relativistic processes that are not representable at the non-relativistic level (such as particle creation and annihilation, and virtual propagation), absorption is quantitatively defined in unambiguous physical terms. In addition, specifics of the relativistic transactional model demonstrate that the Maudlin 'contingent absorber' challenge to the original TI cannot even be mounted: basic features of established relativistic field theories (in particular, the asymmetry between field sources and the bosonic fields, and the fact that slow-moving bound states, such as atoms, are not offer waves) dictate that the 'slow-moving offer wave' required for the challenge scenario cannot exist. It is concluded that issues previously considered obstacles for the Transactional Interpretation are no longer legitimately viewed as such, and that reconsideration of the model is warranted in connection with solving the measurement problem.

1. Introduction and Background

In a nutshell, the measurement problem (MP) is this: given an interaction among quantum systems (such as an unstable atom, atoms comprising a Geiger Counter, atoms comprising a vial of gas, a cat, a friend of Wigner, etc.), which of those interactions constitutes 'measurement,' and why? During the past several decades, worries about the MP largely abated due to a popular sense that environmental decoherence took care of defining measurement in a unitary-only picture (even though there were numerous criticisms of that approach—e.g., Dugić and Jeknić-Dugić, 2012; Fields, 2010; Kastner, 2014c). However, there remains a marked lack of consensus, and recently there has been a resurgence of concern around this issue. Griffiths goes so far as to remark that:

> …the failure of quantum physicists to solve the measurement problem(s) is
> not only an intellectual embarrassment…but also a serious impediment to
> ongoing research in areas such as quantum information, where understanding microscopic quantum properties and how they depend on time is central to the enterprise. (Griffiths, 2017)

However, perhaps the situation is not so dire. The present author would like to issue a gentle reminder that in fact there is a strong contender for solving the measurement problem in the Relativistic Transactional Interpretation (e.g., Kastner, 2012), which extends and elaborates the original TI of Cramer (1986). Making that extension clear is a major objective of the present work. First, however, it is well known that about a decade after Cramer's original proposal, Maudlin (1996; 2nd ed. 2002) raised what appeared at the time to be a fatal objection to TI, and at that point a consensus developed that TI was not viable. What went largely unnoticed after Maudlin's apparent disposal of TI were several publications demonstrating that the Maudlin objection was not in fact fatal (e.g., Marchildon, 2006; Kastner, 2006; Kastner 2012, Chapter 5). More importantly, however, is that the Maudlin objection is itself completely nonviable once the relativistic level of the transactional picture (RTI) is taken into account (Kastner 2017a).

In view of the ongoing concern about the MP, this more recent nullification of the Maudlin objection is briefly reviewed herein, as well as the RTI solution to the measurement problem, including specific, quantitative criteria for the processes of emission and absorption (Kastner 2012, Section 6.3.4). This development does not seem to have penetrated the community, since a recent review by L. Marchildon of Cramer's latest book (Cramer 2016) completely omits it. Based only on the older version of TI presented in Cramer's book, Marchildon expresses his worry that

> "In an important sense, TI is not better defined than the the Copenhagen interpretation...in Cramer's view, transactions play the part of collapse. True, they are somewhat immune to questions like "When does the collapse occur?," but they require emitters and absorbers. These should be macroscopic (classical) objects if transactions are truly irreversible. The classical-quantum distinction or apparatus definition therefore plagues Cramer's view just as it does Bohr's or von Neumann's." (Marchildon 2017)

In fact, however, this is no longer the case. Emission and absorption are now quantitatively defined at the microscopic level, and the microscopic/macroscopic transition is quantitatively defined (although fundamentally indeterministic).[1] So the issue leading to Marchildon's assessment that TI fares no better than the Copenhagen Interpretation[2] is precisely what has been resolved in the relativistic extension of TI (RTI). Since this is a serious misunderstanding of the present status of the transactional interpretation, I shall deal with that first (following a brief review of basic principles of TI), and shall subsequently review the nullification of the Maudlin challenge.

---

[1] Also, Marchildon presupposes that one needs a macroscopic object to provide irreversibility. That this need not be the case, and that RTI explains why, is pointed out in Kastner (2017b).
[2] Although this is perhaps going too far, since TI at least provides a physical account of the form of the Born Rule, lacking in the Copenhagen approach.

2. The basics: a brief review

Cramer's original version of TI (Cramer, 1986) was based on the Wheeler-Feynman direct-action theory of fields (Wheeler and Feynman 1945, 1949). The more recent development by the present author (which resolves the above issues raised by Marchildon) is based on Davies' relativistic extension of the basic W-F theory (Davies 1971, 1972). The direct-action theory has historically been disregarded, since Feynman abandoned it, and this may have led to its unpopularity. However, there is nothing technically wrong with it, and Wheeler himself urged reconsideration of the direct-action picture of fields (Wesley and Wheeler, 2003). See also Kastner (2016c) for an account of how Feynman's abandonment of the direct-action theory had to do with his own particular goals, and expectations, not due to any real defect of the direct-action theory itself;[3] and that it can also serve to remedy the consistency problems afflicting standard quantum theory as expressed in Haag's Theorem (Kastner, 2015).

RTI is introduced in Kastner (2012).[4] For a quick review of the basics, including the concepts of 'offer wave' (OW) and 'confirmation wave' (CW), and the TI derivation of the Born Rule, see Kastner (2016a). For present purposes, an offer wave corresponds to the standard quantum state vector |X>, while a confirmation wave corresponds to the advanced response of an absorber, represented by dual vector <Y|. However, it is important to note that in order to qualify as genuine offer waves, capable of generating confirmations, these states must refer to excitations of single fields—not bound states, such as atoms, where the representation |X> describes a center-of-mass coordinate rather than a 'quantum field of the atom,' of which there is none. This issue will be discussed further below, in connection with the Maudlin challenge.

TI explains, in *physical terms*, what constitutes 'measurement': measurement occurs when there is absorber response (generation of one or more CW). This is a real physical process, albeit an indeterministic one. However, Marchildon raises the concerns: what physically defines emission and absorption? What makes something an 'emitter' or an 'absorber'? What is required for OW and CW to be generated? Pessimism regarding the solubility of these issues is understandable, in view of the seemingly infinite regress encountered in other interpretations (e.g. 'Wigner's Friend'). However, the quantum relativistic level of the direct-action theory does allow a quantitative and well-defined termination of what seemed, based on previous efforts confined to the non-relativistic theory, to be an infinite regress. The relativistic level of RTI is underlain by the Davies quantum relativistic direct-action theory (Davies 1971, 1972). For further background, rather than repeat here what has already been published about the relativistic extension of TI (RTI) and its relation to the Davies theory, the reader is invited to consult Kastner ( 2012b and 2014a ). It may also be helpful to review Kastner (2016a).

---

[3] For example, Feynman wanted a direct-action theory with no self-action, and when he found that some form of self-action was required for relativistic effects such as the Lamb shift, he abandoned it. Kastner (2015, 2016b) discusses why the direct-action theory is still of value when self-action is included. Of course, Davies still viewed the direct-action theory as worthwhile, since he developed it after Feynman abandoned it (Davies 1971, 1972). Thus the existence of self-action is no reason to discard the theory. In fact, self-action (virtual) divergences are 'defanged' in the direct-action theory, since they do not represent the exchange of energies but only of forces.

[4] In Kastner (2012), the extended TI was referred to as the 'possibilist transactional interpretation' or 'PTI,' but I now suggest 'RTI' to emphasize that this is a fully relativistic version that precisely defines emission and absorption.

The above references will hopefully serve to establish that there really is new physical content at the relativistic level that can serve to define 'measurement' and to provide a terminus to what seems like an 'infinite regress' when one considers only nonrelativistic quantum mechanics, which is limited as to what it can describe. In fact, the lesson hopefully to be gained from what follows is that 'measurement' can only be fully and satisfactorily described at the relativistic level. This should perhaps not be terribly surprising, since in order for there to be a measurement, something has to be detected. Detection is fundamentally particle annihilation, but that is always a relativistic process; the nonrelativistic theory only describes persistent particles.

First, I present 'short-answer' versions of the answers to the questions raised in Marchildon (2017), introducing some hopefully helpful terminology; later, I elaborate further.

1. micro-emitter: an excited atom or molecule (i.e. bound state)
2. micro-absorber: a ground-state atom or molecule or one that can be excited further
3. macro-emitter: a collection of N (N>>1) micro-emitters
4. macro-absorber: a collection of N (N>>1) micro-absorbers
5. emission: a micro-emitter emits an OW ($|\omega>$; in general, a very close approximation to a spherical wave of frequency $\omega$)
6. absorber response: a micro-absorber generates a CW corresponding to the component of OW received by it ($<\omega,\mathbf{k}|$). This instantiates the non-unitary measurement transition (von Neumann 'Process 1').
7. absorption: actualized transaction in which real conserved quantities are transferred from the emitter to a particular micro-absorber, resulting in excitation of the latter. *This is irreversible (non-unitary) at the level of the micro-absorber*, so irreversibility does not require a macroscopic absorber in TI (contrary to Marchildon's assumption; see note 1).

Regarding 1-4: Bound states are well-defined in physics, and regarding 5-6: it is already known that these have well-defined (time-dependent) amplitudes for decay and for excitation (transitions between states). For example, the relevant transition amplitudes for the case of atomic electron transitions by way of the emission and absorption of photons are

$$c_m = \frac{1}{i\hbar} \langle m | H_I | l \rangle \int_0^\tau dt \exp[\frac{i}{\hbar}(E_m - E_l \mp \hbar\omega)t] \qquad (1)$$

where *l* and *m* denote the initial and final states respectively, and $H_I$ is the time-independent part of the interaction Hamiltonian (e.g. Sakurai 1973, p.40). These standard transition amplitudes serve to define and quantify emission and absorption in RTI, which is an indeterministic process at the micro-level, as is evident from (1).

Perhaps the reader can already begin to see what we can gain from taking explicitly into account processes described by (1), but we now lay out the key point that remedies lacunae in the 1986 version of TI. First, let us note the explicit form of the interaction Hamiltonian $H_I$ in (1), to see that it carries a factor of the coupling amplitude for quantum electrodynamics (QED), i.e., the elementary charge *e*. Specifically (to lowest order and considering only one atomic electron), we have:[5]

$$H_I = -\frac{e}{mc}(\vec{A}(\vec{x},t) \cdot \vec{p}) \qquad (2)$$

Taking into account the coupling amplitude in the interaction Hamiltonian between the electromagnetic field and its sources (charged fields such as electron currents) is a crucial aspect of the relativistic development, which provides a precise and quantitative answer to the questions above. The crucial development allowing definition of measurement in the relativistic RTI is:

> 8. The coupling amplitude *e (natural units)* is identified as *the amplitude for an offer or confirmation to be generated*.

Note that this is exactly consistent with Feynman's observation, regarding QED, that the coupling amplitude is the amplitude for a charged current to emit or absorb a real photon (Feynman 1985). This generalizes to any form of charge, as in the color charge. That is, charges are just coupling amplitudes: the amplitude for emission of an OW or generation of a CW (where the latter does not necessarily constitute actual absorption of the real quantum, see below). Whether the amplitude at any particular interaction vertex will describe generation of an OW or of a CW is dependent on satisfaction of energy conservation and the relevant selection rules (e.g., a ground state atom cannot emit, and more generally, there can be no OW or CW for a forbidden transition). Provided the relevant transition is not forbidden, the basic probability of emission of an OW or of generation of a CW is given by *square of the coupling amplitude*, since both an emitting and an absorbing current (micro-emitter and absorber respectively) are required in the direct-action picture; i.e., OW emission and CW generation must occur jointly. The square of the coupling amplitude is the fine structure constant *$\alpha$ = 1/137 ~ .007.*

---

[5] Cf. Sakurai (1973), p. 36. Of course, the standard theory (QED) works with a quantized electromagnetic field. For application to the direct-action theory, we must keep in mind that the 'particle-field interaction' where the latter is a quantized field *A* is replaced by a direct connection (time-symmetric propagator) between micro-emitters and micro-absorbers, or 'currents' (see Davies 1971, p. 837). Thus, spontaneous emission in the direct-action picture is due to the presence of absorbers; nothing is emitted without the existence of absorbers.

Another technical detail needs to be made explicit, since in TI, the term 'absorption' can be ambiguous. Micro-absorbers can respond with CW to an emitter, but will not necessarily end up 'winning the competition' to actually absorb the real photon. This is reflected in the fact that the square of the coupling amplitude is the probability for CW generation *only*, while the square of the relevant transition amplitude is the probability (given CW generation) of absorption of a real photon whose properties correspond to that transition amplitude.

Since generation of a CW in response to an OW defines the measurement transition (which could result in a null measurement if the micro-absorber generating the CW does not absorb the real photon), #8 defines *precisely under what physical circumstances, and with what probability, the measurement transition occurs*. The above applies to any micro-emitter E or micro-absorber G. Of course, given only a single micro-absorber, the probability of the measurement transition occurring is very small (.007). We will see this quantitatively below, and then show what is needed to increase that probability to near-certainty, and that this is what defines the macroscopic level.

Thus, we get an unambiguous answer to the question of what precipitates the measurement transition in physical terms, vacating the concern that TI 'does not define emission or absorption.' At this stage (a single micro-emitter E and micro-absorber G), it is an indeterministic account; but that should not surprise us, given that quantum theory (absent the *ad hoc* addition of hidden variables) otherwise has intrinsic objective indeterminacy.

Interestingly, this naturally leads to a criterion for the microscopic vs. macroscopic levels, as follows. We first need to remark that, if more than one micro-absorber is available, we get a 'competition' among all responding absorbers such that only one of them 'wins' and becomes the 'receiving absorber.'[6] However, the non-unitary measurement transition occurs once a confirmation (CW) is generated (see Kastner 2012, Chapter 3 for specifics). It is not required that the micro-absorber that generated the CW actually 'wins' and absorbs the real photon. For any situation in which more than one micro-absorber generates CW, the photon will in fact be absorbed somewhere,[7] and that is all that is necessary for the measurement transition to occur. In fact, and importantly, this is what allows TI to explain 'null measurements': the fact that an absorber generated a CW dictates that a measurement took place, even if the photon is never detected at that absorber.

Now, to the micro/macro distinction provided by RTI. We can understand a macro-absorber as something like a detector, which we can manipulate in the lab, i.e., place in an experimental setup so that an emitted particle can be detected there with virtual certainty (in any given unit time). [8] Well, what is such a detector? It is simply a conglomerate of many micro-absorbers $N$, each one playing the part of G above, such that the probability of *at least one* of the $N$ micro-absorbers generating a CW approaches unity. How big does $N$ need to be for this? It turns out that what we consider 'macroscopic' corresponds very nicely to this criterion. For example, take a sample of metal containing $N$ loosely bound electrons, each capable of being excited via the above sort of interaction between E and the single micro-absorber G. As above, the basic probability of CW generation applying to each of the electrons is the fine-structure constant $\alpha$. But there are $N$ of them now comprising our detector D, and all we need for D to count as a macro-absorber, and therefore as a measuring instrument (in unit time) is for *any one (or at least one) of the N electrons* of D to generate a confirmation wave within the relevant unit time.

---

[6] This is the 'collapse' stage, and proceeds via a generalization of spontaneous symmetry breaking; cf. Kastner 2012, Chapter 4.

[7] Of course, the transition probabilities for decays are time-dependent. So it cannot be precisely specified *when* real absorption will occur.

This is easy to calculate if we first find the probability of the complement: i.e., how likely is it that for $N$ micro-absorbers constituting D, there will be *no* confirming response to micro-emitter E? Let us call this Prob(no CW). For the previous case of a single G,

$$\text{Prob}_{(N=1)}(\text{no CW}) = 1-\alpha = 0.993, \qquad (3)$$

So it's very unlikely that our single G will count as a 'detector,' in that it will very likely not trigger the measurement transition (although it is remotely possible). For N>1, the probability that not a single micro-absorber (electron) constituent of D generates a CW is

$$\text{Prob}_{(N)}(\text{no CW}) \sim (1-\alpha)^N = 0.993^N. \qquad (4)$$

We can see that as N increases, this quantity will decrease. If we consider a small but macroscopic sample of metal, containing about $N=10^{23}$ excitable electrons, we find

$$\text{Prob}_{(N=10^{\wedge}23)}(\text{no CW}) \sim 0.993^{(10^{\wedge}23)} \sim 0. \qquad (5)$$

Thus, given a sample D with $10^{23}$ excitable electrons, *the probability that not one of them will respond to micro-emitter E* is virtually nil. This means that, with virtual certainty, at least one micro-absorber constituent of D will respond, in which case D has responded (since it does not matter which of D's electrons responds). The virtual certainty that D will respond confers upon it the status of 'macro-absorber,' in that *it reliably triggers the measurement transition*. So this is where the buck stops, and why it stops here. This account clearly delineates the micro/macro transition point, as follows:

---

[8] In this analysis, I presuppose the simplest detection situation, i.e., one in which there is an ordinary photon source and a 'blob' of ground state atoms, with no other correlating degrees of freedom, interactions, or filters that would (for example) impede excitation of the atoms from ground state to excited (stationary) states. Obviously, for a more complicated arrangement, the analysis and predictions would differ.

Definition: An object O is a 'macroscopic object' (functioning as an absorber) if at least one of its absorbing constituents (micro-absorbers) responds with CW to an interacting micro-emitter E.

Interestingly, the same analysis allows us to define the 'mesoscopic' level—this is a level involving fairly large and complex systems compared to elementary particles, yet still retaining some quantum features (such as a 'Buckeyball' molecule, comprising 60 carbon atoms). Mesoscopic objects would comprise numbers $N$ of micro-absorbers such that they would have a significant but still uncertain probability of CW response to an emitter. Let us suppose, just as a crude estimate, that the Buckeyball's 60 carbon atoms correspond to 60 excitable degrees of freedom (micro-absorbers). This gives us a value for Prob(no CW) of:

$$\text{Prob}_{N=60}(\text{no CW}) \sim .993^{60} \sim .66 \qquad (6)$$

Thus, it is quite possible that a Buckeyball will respond with a CW (i.e., Prob (CW) = 1–Prob(noCW) =.34), but far from certain. In this way, it can be seen that the probability of CW generation by any given object, based on the number $N$ of its constituent micro-absorbers, provides a clear physical and quantitative criterion for whether that object qualifies as 'macroscopic' (meaning virtual certainty that it precipitates the non-unitary measurement transition and therefore qualifies as a detector, or basic measurement apparatus) , 'microscopic' (extremely unlikely to precipitate the transition) or 'mesoscopic' (somewhat likely to precipitate the transition).

The above result, based upon the quantitative criterion for the measurement transition (i.e. the coupling amplitude interpreted as amplitude for OW or CW generation), is probably the most important of the developments of RTI. It demonstrates that RTI remedies lacunae in the original TI, in which emitters and absorbers were essentially primitive notions. If the present author is not mistaken, this addresses the notorious problem of the 'Heisenberg Cut' between unitary evolution and the non-unitary von Neumann 'Process 1' instantiating the Born Rule. As noted above, it is not a 'cut' so much as a range of values of $N$ (number of constituents of any particular object) in which the measurement transition at that object becomes more and more likely until it is virtually assured. The latter means the object is 'macroscopic' and an absorber (or mutatis mutandis for an emitter) in that it generates CW (or emits OW) with certainty in a relevant unit time.

3. Nullification of the Maudlin Challenge

Maudlin's challenge (e.g., Maudlin 2002) was a worthy one, in that it spurred further development of TI, both into the relativistic realm and in ontological terms. The challenge is reviewed and refuted in Kastner 2017a. It is a thought experiment in which a purported 'slow-moving offer wave' is emitted in a superposition of rightward and leftward momenta. To the right is a detector A, at a distance $x$. There is no detector initially on the left (although, as pointed out in Marchildon, 2006, there is always some background absorber C if anything is taken as being 'emitted' to the left; this was one of the earlier refutations of Maudlin's objection).

Behind detector A on the right, at a distance of $2x$ from the source, is a detector B which is moveable so that it can be swung around very quickly to the left as required. Maudlin assumes that the time of arrival of the rightward component of the 'slow-moving offer wave' is well-defined, and after that time has passed, if there is no detection at A, B is swung quickly around to the left to intercept the quantum. Maudlin worried that the probability of 'detection at B' is only ½ according to the OW/CW interaction, though whenever it is detected there, its detection is certain. He viewed this as an inconsistency. However, there is no inconsistency, since the probability of ½ need not be defined as 'detection at B' but rather can be understood as 'detection of a quantum with leftward momentum,' which occurs precisely half the time in the experiment (as noted by Marchildon, there is always a CW from the left, whether or not B is swung around). And in fact the observable being measured in the experiment *is* directional momentum, not 'which detector, A or B'. This is because the quantum has been prepared in a superposition of momenta, not a superposition of 'detectors A and B').

Maudlin was also concerned that Cramer's 'pseudotime' account could not deal appropriately with situations like this, in which the existence of a CW is contingent on an earlier detection result. The present author has addressed these sorts of situations—contingent absorber experiments or CAE—in Kastner 2012, Chapter 5, showing that to the extent they are possible, they raise perplexing issues not just for TI but for standard quantum mechanics.

Nevertheless, it turns out that in fact there is no 'slow-moving offer wave' as is required for Maudlin's challenge to be mounted. This becomes evident at the relativistic level in RTI. Specific details are provided in Kastner (2017a). For our present purposes, we can summarize as follows: in order to have a 'slow-moving offer wave', one must be working with a massive object. Such objects are either bound states (e.g. atoms), which are not offer waves (see Kastner, 2017a for why); or they are excitations of matter fields that are sources of boson fields (such as an electron, which is a source of photons). Regarding the latter, transactions involving fermions are always mediated by bosons (force carriers). There is no 'transaction' involving only a 'fermion offer wave' and a 'fermion confirmation wave'. This is due to the intrinsic asymmetry between fields and their sources--technical specifics are provided in Kastner (2017a). So there simply is no 'slow-moving offer wave' that could be subject to the kind of contingent absorption proposed by Maudlin. Such a situation does not exist in physics. If the quantum is slow-moving, it is a bound state, not an offer wave; or (assuming this were possible) a slow electron which is never confirmed by an 'electron confirmation wave' but only indirectly via photon transactions. (Massive bosons would be of no use since they are always short-range.)

The reader may still be worried about the idea of an 'orphan' offer wave being emitted to the left with no absorber present on that side at all (even though there can be no 'slow-moving offer wave' as above). However, once the relativistic level is taken into account, it is clear that there can never be any offer wave (more precisely, offer wave component) emitted without absorber participation. This is the quantum relativistic analog of the Wheeler-Feynman 'light tight box' condition, except that it is no longer *ad hoc* but is an intrinsic part of the dynamics of the direct-action theory, as follows. Both emitter and absorber contribute mutually to the elevation of a virtual (time-symmetric propagator) connection between them, which lacks any temporal orientation, to a real photon (time-asymmetric field corresponding to a projection operator, where the ket or Fock state component of that operator is the offer wave); see Kastner 2014a. Put differently, both emitter and absorber are necessary but not sufficient conditions for OW and CW generation, respectively (where the lack of sufficiency is simply the fact that OW and CW generation are subject to fundamental indeterminacy, reflected in the coupling amplitude, and thus not assured). Thus, at the quantum relativistic level of the Davies theory, it is seen that absorber response is a minimum requirement for any offer wave or offer wave component, so that if there is no absorber for that component, no such offer wave component exists. The distinction between virtual photons (time-symmetric propagator, no absorber response) and real photons (pole in the Feynman propagator, established through absorber response) is discussed in Kastner (2014a). Further technical details are discussed in Kastner and Cramer (2017).

Perhaps another way to understand this analog of the 'light-tight box' condition is as follows: in the original Wheeler-Feynman theory, the condition was presented as "all emitted fields must be absorbed." Instead, in the Davies theory and in RTI, the condition is "no field (meaning real field or Fock state) is emitted unless there is absorber response." Physically, this means that the emitter and absorber mutually create the emitted field; both are required. This is the essence of the direct-action theory.

4. Ontological considerations

In a private communication, Maudlin (2017)[9] worried that there is no real collapse in TI because the time-symmetric character appears to demand that all events (including absorber responses) already exist in a static block world; so there is no real dynamics, including no real collapse. I agree with this concern, which applies only to the original 1986 TI. In fact I have argued that the same problem applies to all 'time-symmetric' interpretations that contain explicit or implicit future boundary conditions (Kastner 2017c). RTI, however, proposes an expanded ontology in which OW and CW are pre-spacetime processes, along the lines of Heisenberg's 'potentiae' (see Kastner, Kauffman and Epperson, 2017.)

---

[9] Email from T. Maudlin to J. Gibson, provided to the author.

In RTI, spacetime is considered a relational (non-substantival) causal set structure emergent from a quantum substratum, such that spacetime events are added to the causal set with every actualized transaction. For details, see Kastner, 2014b. In particular, micro-emitters and micro-absorbers are elements of the quantum substratum, not spacetime objects, and this is what permits escape from the static block world ontology. In this picture, spacetime emergence is a dynamical process, in which 'collapse' is the establishment of a spacetime interval connecting an emission and absorption event. The resulting spacetime causal set grows time-asymmetrically, since all absorption events are to the future of their emission events. There is no future boundary condition; the future is genuinely open. Thus, RTI contains real dynamics, and real collapse. For additional details regarding the breaking of time symmetry at the level of spacetime in RTI, see Kastner 2017b, Section 4. Thus, once again, Maudlin's concern is with the original TI, not the extended and re-formulated relativistic version that is RTI.

In case the proposed expanded ontology, renouncing spacetime as the arbiter of what can be considered 'physically real' seems too outlandish, it is worthwhile recalling Zeilinger's recent insightful observation:

> ..it appears that on the level of measurements of properties of members of an entangled ensemble, quantum physics is oblivious to space and time… It remains to be seen what the consequences are for our notions of space and time, or space-time for that matter. Space-time itself cannot be above or beyond such considerations. I suggest we need a new deep analysis of space-time, a conceptual analysis maybe analogous to the one done by the Viennese physicist-philosopher Ernst Mach who kicked Newton's absolute space and absolute time form their throne. (Zeilinger 2016)

5. Conclusion.

It has been argued that once the relativistic developments of the transactional interpretation (RTI) are taken into account, the transactional picture does in fact solve the measurement problem by clearly defining 'emitters' and 'absorbers' and specifying the quantitative physical circumstances that trigger the non-unitary measurement transition. Moreover, this development allows a natural account of the micro/meso/macro transition zones, allowing us to understand why, for example, a single electron cannot serve as a 'measurement apparatus,' while many of them (suitably bound) can do so. Thus, the foregoing resolves the notorious conceptual problem of the 'Heisenberg Cut.' It provides an objective, physical account of 'measurement' from within the theory, without any need to refer to an 'external consciousness,' a notoriously ill-defined concept.

Author's Postscript.

While it is obvious that I have been an advocate of the Transactional Interpretation since 2012, I am not interested in pursuing any interpretation that does not work. Thus, I am not a 'true believer' in TI; if presented with a critical flaw in the proposed interpretation, I would be happy to waste no further time and effort on it. However, to date, I have not seen any substantive objection to the 2012 version, RTI (although I continue to see objections to the 1986 version; but that version has been supplanted in the peer-reviewed literature with developments fully addressing those objections). As reviewed herein, the Maudlin objection fails completely at the relativistic level of the Transactional Interpretation (RTI). Absorbers and emitters are now precisely and quantitatively defined, along with the specific physical conditions precipitating the measurement transition. Thus, it seems important to continue to explore the model, which appears to have stood the test of time and weathered all objections of which I am aware.[10] I am certainly open to hearing new concerns that in the view of the reader have not been sufficiently addressed. Again, if the model were at some point shown to be nonviable, I would have no further interest in it. But what would need to be addressed is the 2012 RTI (Kastner 2012 and as clarified herein), not the 1986 TI.

Acknowledgments. The author is grateful to N. Gisin for valuable correspondence, and to R. Scheffer for a careful reading of the manuscript.

---

[10] In a private communication, N. Gisin raised the possibility that an experiment discussed in Staudt et al (2007) was a 'falsification' of this proposal. However, that experiment involves creating a superposition of ground and excited states of atoms described as briefly 'storing' a coherent state of the field (as opposed to a Fock state with definite photon number). Under these conditions, absorption has not taken place at the level of any of those atoms (they have not transitioned from ground to excited state). Thus, the usage 'storing a field' in the description of the experiment of Staudt et al (2007) does not equate to 'absorbing a photon.' A photon is a Fock state, and its absorption results in excitation (to a stationary state) of the system receiving it, with an accompanying decrement of the occupation number of the field (as discussed in Sakurai, 1973; p.37). That does not obtain in the above experiment, which employs very special conditions to maintain coherence of the fields being temporarily stored—conditions not applying to the situation discussed in eqs. 3-6 herein. Moreover, RTI is empirically equivalent to standard quantum theory, and cannot be 'falsified' by experiments conforming to the predictions of standard quantum theory. The only sense in which RTI departs from standard quantum theory is to explain, rather than to assume, the measurement transition.